\documentclass[aps,prl,groupedaddress,showkeys,preprint]{revtex4}
\usepackage{graphicx}
\usepackage[dvips]{epsfig}
\usepackage{dcolumn}
\usepackage{subfigure}%
\usepackage{bm}
\usepackage{amssymb}
\usepackage{amsmath}

\begin{document}
\title{Blow-up of the hyperbolic Burgers equation}

\author{Carlos Escudero\footnote{Mathematical Institute, University of Oxford, 24-29 St Giles', Oxford OX1 3LB, United Kingdom. Tel.: +44 (0) 1865 283891. Fax: +44 (0) 1865 273583. E-mail address: escudero@maths.ox.ac.uk.}}

\noaffiliation

\begin{abstract}
The memory effects on microscopic kinetic systems have been sometimes modelled by means of the introduction of second order time derivatives in the macroscopic hydrodynamic equations. One prototypical example is the hyperbolic modification of the
Burgers equation, that has been introduced to clarify the interplay of hyperbolicity and nonlinear hydrodynamic evolution. Previous studies suggested the finite time blow-up of this equation, and here we present a rigorous proof of this fact.
\end{abstract}

\maketitle

One of the major problems in nonequilibrium statistical mechanics is to derive an accurate hydrodynamic description of a kinetic process. A classical example is the derivation of the Euler and Navier-Stokes equations from the Boltzmann
equation~\cite{golse,raymond}, for instance by means of the Chapman-Enskog expansion~\cite{chapman,cercignani}. While the Navier-Stokes equations represent a remarkable success of the theoretical study of fluid mechanics, it is well known that the spectral properties of their solution do not agree with experimental data for short wavelengths~\cite{chen}. In order to solve this problem, one is tempted to derive generalized hydrodynamic equations continuing the Chapman-Enskog
expansion to higher orders~\cite{burnett}, to obtain the so called Burnett equations, but they have never achieved any notable success~\cite{grad}. Another possibility was suggested by Rosenau~\cite{rosenau}, who found that the telegraphers equation
\begin{equation}
\partial_{tt}u+\partial_tu=\partial_{xx}u
\end{equation}
reproduces the spectrum of its microscopic counterpart, the persisting random walk, almost exactly. Using this fact, he claimed that the Chapman-Enskog expansion should be substituted by a different expansion keeping space and time on equal footing. This procedure would preserve the hyperbolic nature of the resulting equations, and thus the nice spectral properties of the solution, at least in the linear regime~\cite{rosenau}. An expansion of this type was carried out by Khonkin~\cite{khonkin}, who found new equations for the momentum and energy fluxes which, in contrast to the classical Navier-Stokes and Fourier laws, depend on the first time derivative of these fluxes. This dependence implies in turn the appearence of a term proportional to the second order time derivative of the velocity, among others, in the corresponding modified Navier-Stokes equation, which becomes now hyperbolic. However, it was already argued by Rosenau that hyperbolicity in union with nonlinear hydrodynamical evolution might result in the nonexistence of the solution. A similar idea was proposed later~\cite{makarenko}, where hyperbolicity was introduced to take into account memory effects in the hydrodynamic description of the flow, and to get rid of the infinite speed of signal propagation. In order to understand the interplay between hyperbolicity and nonlinear convection, a hyperbolic modification of the Burgers equation
(see Eq. (\ref{hburgers}) below) was studied in Ref.~\cite{makarenko} by means of linear and numerical analyses. One of the conclusions of this work is that this equation has blowing up solutions under certain circumstances. Surprisingly, there is not much rigorous work done with respect to this equation. To our knowledge, there is only one result proving global existence in time of the solution provided the initial data is small enough~\cite{rudiak}. It is thus our goal to extend this result and prove blow-up of the solution for large initial data. But before starting with the analysis, let us discuss the validity of the hyperbolic Burgers equation as a physical model. It is important to note that the hyperbolic hydrodynamic equations derived from the Boltzmann equation~\cite{khonkin} and obtained including memory 
effects~\cite{makarenko} are far more complex than the equation under study. Furthermore, there is, to our knowledge, no direct application of the hyperbolic Burgers equation to a physical problem. It is thus to be interpreted as a model equation resembling some of the characteristic properties of the full models, which will hopefully help us to understand the dynamical properties of the more difficult equations.

In this article we are concerned with the Cauchy problem for the hyperbolic Burgers equation
\begin{eqnarray}
\label{hburgers}
\mu \partial_{tt} v + \partial_t v + v \partial_x v = \nu \partial_{xx}v, \\
v(x,0)=v_0(x), \\
v_t(x,0)=v_1(x).
\end{eqnarray}
where $v$ represents the one-dimensional velocity and $\mu>0$ and $\nu>0$ stand for the inertia and the viscosity respectively. From now on we will assume that the initial conditions are compactly supported in the interval $I_0=[-L,L]$, for some $0<L<\infty$. We will use the notation $c=\sqrt{\nu/\mu}$, because, as we will see, this quantity will play the role of the speed of sound in Eq.~(\ref{hburgers}), that is, the maximum velocity of propagation of disturbances. The proof presents some similarities to the corresponding one of nonexistence of the solution to the compressible Euler
equations~\cite{sideris}, but, interestingly, the singular behaviour developed by the hyperbolic Burgers flow is not of the
shock-wave type. Actually, what diverges in this case is the velocity itself, as we will show.

To prove the finite time blow-up of Eq.(\ref{hburgers}) we need first some result concerning the finite velocity of propagation of disturbances.

{\bf Proposition.} {\it Let C(x,t) be the cone}
\begin{equation}
C(x,t)=\{(x,t): |x_c-x| \le c(t_c-t)\}, \qquad 0 \le t \le t_c,
\end{equation}
{\it for some $x_c \in \mathbb{R}$, $t_c \in \mathbb{R}^+$.
If $v$ is a classical solution of Eq.(\ref{hburgers}) in $C(x,t)$ and $v_0$, $v_1 \equiv 0$ in $C(x,0)$, then $v \equiv 0$ in
$C(x,t)$, $0 \le t \le t_c$.}

{\bf Proof.} Let us rewrite Eq.(\ref{hburgers}) in the form
\begin{equation}
\partial_{tt} v + \frac{1}{\mu}\partial_t v + \frac{1}{\mu} v \partial_x v = \frac{\nu}{\mu} \partial_{xx}v,
\end{equation}
and define the energy functional
\begin{equation}
E(t)=\frac{1}{2}\int_C (v_t^2 + c^2 v_x^2)dx.
\end{equation}
The first derivative of the energy functional yields
\begin{equation}
E'(t)=\int_C (v_t v_{tt} + c^2 v_x v_{xt})dx-\frac{c}{2}[v_t^2 + c^2 v_x^2]_{\partial C^-}-\frac{c}{2}[v_t^2 + c^2 v_x^2]_{\partial C^+},
\end{equation}
where $\partial C^{+(-)}$ denotes the right (left) boundary of $C(x,t)$ for a fixed time, and an expression between square brackets with this subindex indicates that it is evaluated at this point. We can further proceed integrating by parts
\begin{equation}
E'(t)=\int_C v_t(v_{tt}-c^2 v_{xx})dx +c^2 [v_x v_t]_{\partial C^+}-c^2 [v_x v_t]_{\partial C^-}-\frac{c}{2}[v_t^2 + c^2 v_x^2]_{\partial C^-}-\frac{c}{2}[v_t^2 + c^2 v_x^2]_{\partial C^+},
\end{equation}
and using Cauchy inequality in the boundary terms
\begin{equation}
c^2 [v_x v_t]_{\partial C^+}-c^2 [v_x v_t]_{\partial C^-} \le \frac{c}{2} \left( [v_t^2 + c^2 v_x^2]_{\partial C^+}
+ [v_t^2 + c^2 v_x^2]_{\partial C^-} \right),
\end{equation}
in order to get
\begin{eqnarray}
\nonumber
E'(t) \le \int_C v_t(v_{tt}-c^2 v_{xx})dx= -\frac{1}{\mu}\int_C v_t^2 dx -\frac{1}{\mu}\int_C v_t v v_x dx \\
\le \frac{1}{\mu} \int_C |v_t v v_x|dx \le \frac{||v||_{L^\infty(0,t_c;L^\infty(C))}}{2 \mu c} \int_C (v_t^2+c^2 v_x^2)dx,
\end{eqnarray}
where we have used Cauchy inequality together with the regularity of classical solutions to Eq.~(\ref{hburgers}). We can integrate this last expression to obtain the inequality
\begin{equation}
E(t) \le \exp \left( \frac{||v||_{L^\infty(0,t_c;L^\infty(C))}}{\mu c} t \right) E(0),
\end{equation}
which, taking into account the fact that $E(0)=0$, yields the identities $v_t$, $v_x \equiv 0$ in $C(x,t)$, and thus, in addition to the initial conditions, the desired result $v \equiv 0$ inside the cone $C(x,t)$. $\Box$

{\bf Remark.} {\it Note that this result also holds for $v$ being a bounded weak solution of Eq.(\ref{hburgers}) in
$C(x,t)$.}

Let us now define $J \subset \mathbb{R}$ as
\begin{equation}
J(t)=\{x \in \mathbb{R}:|x| \le L+ct \}.
\end{equation}
As a consequence of the proposition we know that if our initial conditions are compactly supported in the interval $I_0$, then a classical solution to our problem will be compactly supported in $J(t)$. The proof of our main result will rely on constructing a differential inequality for the quantity
\begin{equation}
F(t)=\int_\mathbb{R}xv dx,
\end{equation}
which is a measure of the expansion of the velocity. We know that this quantity is convergent as long as the solution is regular enough, by direct application of the proposition.

{\bf Theorem 1.} {\it Suppose $v$ is a classical solution of Eq.(\ref{hburgers}) for $0 \le t \le T$. If the initial conditions fulfill}
\begin{subequations}
\begin{eqnarray}
\label{initial1}
F(0)> \frac{16}{3}cL(L+6c\mu), \\
\label{initial2}
F'(0)>\frac{64}{3}c^2(L+6c\mu),
\end{eqnarray}
\end{subequations}
{\it then the life span of the classical solution is finite.}

{\bf Proof.} Consider the effect of the time evolution on $F(t)$:
\begin{equation}
\mu F''(t)+F'(t)=\int_\mathbb{R} x (\mu \partial_{tt}v+\partial_tv)dx=\int_\mathbb{R} x(-v\partial_x v +\nu \partial_{xx}v)dx=\frac{1}{2}\int_\mathbb{R}v^2dx,
\end{equation}
where we have used integration by parts and the compact support property of the classical solutions to Eq.(\ref{hburgers}). Now we derive
\begin{eqnarray}
\nonumber
\left(\int_\mathbb{R} xv dx \right)^2=\left(\int_{I(t)} xv dx \right)^2 \le \int_{I(t)} x^2 dx \int_{I(t)}v^2 dx= \\
\frac{2}{3} (L+ct)^3 \int_{I(t)} v^2 dx = \frac{2}{3} (L+ct)^3 \int_{\mathbb{R}} v^2 dx,
\end{eqnarray}
where we have used again the fact that the solution is compactly supported together with Schwartz inequality. By using these two relations we obtain the following differential inequality
\begin{equation}
\label{ineq}
\mu F''(t)+F'(t) \ge \frac{3}{4}(L+ct)^{-3}F(t)^2.
\end{equation}
We can now adapt the argument in Ref.~\cite{todorova} to prove the blow-up of $F(t)$. Consider the auxiliary initial value problem
\begin{eqnarray}
\label{aux}
G'(t)=\epsilon (ct+L)^{-3}G(t)^{3/2}, \\
G(0)=F(0),
\end{eqnarray}
where $\epsilon>0$ is to be set later. It can be solved to yield
\begin{equation}
G(t)^{-1/2}=G(0)^{-1/2}+\frac{\epsilon}{4c^3} \left[(t+L/c)^{-2}-(L/c)^{-2} \right].
\end{equation}
This solution blows up in a finite time $T^*$ provided the initial condition fulfills
\begin{equation}
\label{cond1}
G(0)>\frac{16c^2L^4}{\epsilon^2},
\end{equation}
and in this case it also increases strictly and monotically from the initial condition to infinity.
Apply now a time derivative on Eq.(\ref{aux}) to find
\begin{equation}
G''(t)=\frac{3}{2}\epsilon^2(ct+L)^{-6}G(t)^2-3\epsilon c(ct+L)^{-4}G(t)^{3/2}\le \frac{3}{2}\epsilon^2(ct+L)^{-6}G(t)^2.
\end{equation}
Summing this last equation multiplied by $\mu$ and Eq.(\ref{aux}) we get
\begin{equation}
\mu G''(t)+G'(t)\le \frac{3}{2}\mu \epsilon^2(ct+L)^{-6}G(t)^2 + \epsilon (ct+L)^{-3}G(t)^{3/2},
\end{equation}
that yields in turn the inequality
\begin{equation}
\mu G''(t)+G'(t) \le \left[\epsilon G(0)^{-1/2}+\frac{3}{2}\frac{\mu}{L^3} \epsilon^2 \right](ct+L)^{-3}G(t)^2.
\end{equation}
Now choose $\epsilon$ small enough so that the relations
\begin{eqnarray}
\label{cond2}
\epsilon G(0)^{-1/2}+\frac{3}{2} \frac{\mu}{L^3} \epsilon^2 \le \frac{3}{4}, \\
G'(0)=\epsilon L^{-3}G(0)^{3/2}<F'(0),
\label{cond3}
\end{eqnarray}
hold. So we have finally reduced our problem to the inequality
\begin{equation}
\label{aux2}
\mu G''(t)+G'(t) \le \frac{3}{4}(ct+L)^{-3}G(t)^2,
\end{equation}
with initial conditions fulfilling $G(0)=F(0)$ and $G'(0) < F'(0)$. Now, it only rests to prove that $F(t) \ge G(t)$. The fact that $F'(0) > G'(0)$ implies that there exists some time $t_0>0$ such that $F'(t)>G'(t)$ if $0 \le t < t_0$. Let us denote by $t_1$ the supremum of all the values of $t_0$. Suppose $t_1<T^*$, then $F(t)>G(t)$ for $0<t<t_1$, and $F'(t_1)=G'(t_1)$. Thus we have $F'(t)-G'(t)>0$ implying that $F(t)-G(t)>F(0)-G(0)=0$ for $0<t<t_1$. This, in turn, implies that $F(t_1)>G(t_1)$, while the difference between Eqs. (\ref{ineq}) and (\ref{aux2}) yields
\begin{equation}
\mu \left[F''(t)-G''(t) \right]+F'(t)-G'(t) \ge \frac{3}{4}(ct+L)^{-3} \left[ F(t)^2-G(t)^2 \right] \ge 0,
\end{equation}
for $0 \le t \le t_1$. This inequality is equivalent to
\begin{equation}
\frac{d}{dt}\left(e^{t/\mu} \left[ F'(t)-G'(t) \right] \right)\ge 0,
\end{equation}
which can be straightforwardly integrated in order to get
\begin{equation}
e^{t_1/\mu} \left[ F'(t_1)-G'(t_1) \right] \ge F'(0)-G'(0),
\end{equation}
or, what is the same, $F'(t_1)>G'(t_1)$. So now we can argue by contradiction that $t_1 \ge T^*$ to conclude the
proof. $\Box$

{\bf Remark.} {\it Note that the requirements on the initial conditions Eqs. (\ref{initial1}) and (\ref{initial2}) come
from Eqs. (\ref{cond1}), (\ref{cond2}) and (\ref{cond3}) in the proof.}

These arguments suggest that the velocity is blowing up in finite time, but they do not constitute a proof of blow-up, just of nonexistence of the classical solution. This is because the maximal time of existence of the classical solution might be strictly smaller than the blow-up time, as happens in certain situations~\cite{ball}. To be sure that the velocity increases unboundedly in finite time we need an additional theorem stating that the solution can be continued as long as we have controlled the $L^\infty$ norm of $v$. For this we can assume the existence of a weak solution to Eq.(\ref{hburgers}); interpreting all the derivatives in the weak sense leaves us with the requirement $v \in L^1 \cap L^2 ([0,T] \times \mathbb{R})$ for the weak solution to exist. This fact, in addition to the proposition, implies that the nonexistence of the weak solution is necessarily due to a divergence of the velocity. Before starting, let us define the set
$\Omega=(0,T) \times U(t) \subset \mathbb{R}^2$, where $U(t)$ is open, bounded, $J(t) \subset U(t) \subset \mathbb{R}$, and such that $\Omega$ has a $C^1$ boundary. Henceforth we will assume that the initial conditions are compactly supported in the interval $I_0$ and they fulfill $v_0 \in H^3(\mathbb{R})$ and $v_1 \in H^2(\mathbb{R})$.

{\bf Theorem 2.} {\it Suppose there exists a weak solution $v$ to the Cauchy problem for Eq.(\ref{hburgers}) for $0 \le t \le T < \infty$. If this solution fulfills $||v||_{L^\infty(\Omega)}<\infty$, then it is a classical solution.}

{\bf Proof.} First of all, we know that a bounded weak solution of Eq.(\ref{hburgers}) has compact support as a consequence of the proposition. Then, it is very easy to see that
\begin{equation}
||v||_{L^2(\Omega)}=\int_0^T \int_{\mathbb{R}} v^2 dx dt \le 2 ||v||_{L^\infty(\Omega)}^2 \left( \frac{c}{2}T^2+LT \right).
\end{equation}
We can now evaluate the quantity
\begin{equation}
E_1(t)=\frac{1}{2}\int_\mathbb{R} (v_t^2 + c^2 v_x^2) dx,
\end{equation}
following a parallel reasoning to that of the proposition, to get
\begin{equation}
E_1(t) \le \exp \left( \frac{||v||_{L^\infty(\Omega)}}{\mu c}t \right)E_1(0).
\end{equation}
This allows us to conclude
\begin{equation}
\int_0^T \int_\mathbb{R} (v_t^2 + c^2 v_x^2) dx dt \le E(0) \frac{\mu c}{||v||_{L^\infty(\Omega)}}
\left[ \exp \left( \frac{||v||_{L^\infty(\Omega)}}{\mu c}T \right) -1 \right].
\end{equation}
Consider now the functional
\begin{equation}
E_2(t)=\frac{1}{2}\int_\mathbb{R} (v_{tt}^2+c^4 v_{xx}^2) dx.
\end{equation}
Its first derivative is
\begin{equation}
\label{e2p}
E_2'(t)=\int_\mathbb{R} (v_{tt}v_{ttt}+c^4 v_{xx}v_{xxt})dx=\int_\mathbb{R} (v_{tt}v_{ttt}-c^4 v_{xxx}v_{xt})dx,
\end{equation}
after derivation by parts. Now we will proceed to evaluate the second integral in the right hand side of the equation above
\begin{equation}
c^2\int_\mathbb{R} v_{xxx}v_{xt} dx= \int_{\mathbb{R}} v_{xt} \left[ v_{ttx} +\frac{1}{\mu}v_{xt}+\frac{1}{\mu}(vv_x)_x \right]=
I_1+I_2+I_3.
\end{equation}
We have
\begin{equation}
I_1=\int_\mathbb{R} v_{xt}v_{xtt}dx =\frac{1}{2} \frac{d}{dt}\int_\mathbb{R} v_{xt}^2,
\end{equation}
\begin{equation}
I_2=\frac{1}{\mu}\int_\mathbb{R}v_{xt}^2 \ge 0,
\end{equation}
and
\begin{equation}
I_3=\frac{1}{\mu}\int_\mathbb{R} (v_{xt}v_x^2 + v_{xt}vv_{xx}) dx \le \frac{1}{3\mu}\frac{d}{dt}\int_\mathbb{R}v_x^3 dx +
\frac{||v||_{L^\infty(\Omega)}}{2 \mu c}\int_\mathbb{R}(v_{xt}^2 +c^2 v_{xx}^2)dx.
\end{equation}
The first integral in the right hand side of Eq.(\ref{e2p}) can be estimated as follows
\begin{equation}
\int_\mathbb{R} v_{tt}v_{ttt} dx = \int_\mathbb{R} v_{tt} \left[ -\frac{1}{\mu}v_{tt}-\frac{1}{\mu}(vv_x)_t+c^2 v_{xxt} \right]dx
=J_1+J_2+J_3.
\end{equation}
We have
\begin{equation}
J_1=-\frac{1}{\mu}\int_\mathbb{R}v_{tt}^2 dx \le 0,
\end{equation}
\begin{equation}
J_2=-\frac{1}{\mu}\int_\mathbb{R}(v_{tt}v_tv_x+v_{tt}vv_{xt})dx \le \frac{||v||_{L^\infty(\Omega)}}{\mu c}
\int_\mathbb{R}(v_{tt}^2+c^2 v_{xt}^2)dx-\frac{1}{2\mu}\int_\mathbb{R}(v_t^2)_tv_x dx,
\end{equation}
and
\begin{equation}
J_3=c^2\int_\mathbb{R} v_{tt}v_{xxt}dx =-c^2 \int_\mathbb{R} v_{xt}v_{xtt}dx=
-\frac{c^2}{2}\frac{d}{dt}\int_\mathbb{R}v_{xt}^2dx.
\end{equation}
Now we can integrate Eq.(\ref{e2p}) with respect to time and use the above expressions to obtain
\begin{eqnarray}
\nonumber
\frac{1}{2}\left[\int_\mathbb{R}(v_{tt}^2+c^4 v_{xx}^2+2c^2 v_{xt}^2) dx \right]_0^{T_1} \le
\frac{c ||v||_{L^\infty(\Omega)}}{2\mu}
\int_0^{T_1} \int_{\mathbb{R}}(v_{xt}^2+c^2 v_{xx}^2)dx dt \\
\nonumber
-\frac{c^2}{3\mu}\left[ \int_\mathbb{R} v_x^3 dx \right]_0^{T_1}
+\frac{||v||_{L^\infty(\Omega)}}{\mu c}\int_0^{T_1} \int_\mathbb{R}(v_{tt}^2+c^2 v_{xt}^2)dx dt \\
+\frac{1}{2\mu}\int_0^{T_1}\int_\mathbb{R}v_t^2v_{xt} dx dt-\frac{1}{2\mu}\left[ \int_\mathbb{R}v_t^2v_x dx \right]_0^{T_1},
\label{e2p2}
\end{eqnarray}
after integration by parts of the last term. We still have to estimate the following terms
\begin{equation}
\int_0^{T_1} \int_\mathbb{R} v_t^2 v_{xt} dx dt=\frac{1}{3}\int_0^{T_1}\int_\mathbb{R}(v_t^3)_x dx dt=0,
\end{equation}
due to the compact support property,
\begin{equation}
\frac{1}{3\mu}\int_\mathbb{R}v_x^3 dx= -\frac{2}{3\mu}\int_\mathbb{R}vv_xv_{xx} dx \le
\frac{2||v||_{L^\infty(\Omega)}}{3\mu}\int_\mathbb{R} \left( \frac{1}{2\epsilon}v_x^2+\frac{\epsilon}{2}v_{xx}^2 \right)dx,
\end{equation}
for $\epsilon>0$, and
\begin{equation}
\frac{1}{2\mu}\int_\mathbb{R}v_t^2v_x dx=-\frac{1}{\mu}\int_\mathbb{R}vv_tv_{xt} dx \le
\frac{||v||_{L^\infty(\Omega)}}{\mu}\int_\mathbb{R}\left( \frac{1}{2\delta}v_t^2 + \frac{\delta}{2}v_{xt}^2 \right) dx.
\end{equation}
for $\delta>0$. Now we can choose $\epsilon$ and $\delta$ small enough, say
\begin{equation}
\epsilon < \frac{3 \mu c^2}{2||v||_{L^\infty(\Omega)}}, \qquad \delta < \frac{2 \mu c^2}{||v||_{L^\infty(\Omega)}},
\end{equation}
and noting that for an arbitrary $T_1$, such that $0 \le T_1 \le T$, we have $||E_1(t)||_{L^\infty(0,T_1)}<C$ for some suitable constant $C$, we can rearrange inequality (\ref{e2p2}) in the following way
\begin{equation}
\left[ \int_\mathbb{R}(v_{tt}^2+c^2 v_{xt}^2+c^4 v_{xx}^2) dx \right](T_1) \le C_1+
C_2 \int_0^{T_1}\int_\mathbb{R}(v_{tt}^2+c^2 v_{xt}^2+c^4 v_{xx}^2) dx dt,
\end{equation}
for some suitable positive constants $C_1$ and $C_2$, and recall Gr\"{o}nwall inequality to find
\begin{equation}
\left[ \int_\mathbb{R}(v_{tt}^2+c^2 v_{xt}^2+c^4 v_{xx}^2) dx \right](T_1) \le C_1 \exp (C_2 T_1).
\end{equation}
Finally, integrating with respect to time yields
\begin{equation}
\int_0^T \int_\mathbb{R}(v_{tt}^2+c^2 v_{xt}^2+c^4 v_{xx}^2) dx dt \le (C_1/C_2)[\exp(C_2 T)-1].
\end{equation}
Now we know that the Sobolev norm
\begin{equation}
\label{sobnorm}
||v||_{H^2(\Omega)}= \left[\int_0^T \int_\mathbb{R} \{(v_{tt}^2+c^2 v_{xt}^2+c^4 v_{xx}^2)\mu^4+(v_t^2+c^2 v_x^2)\mu^2+v^2 \} dx dt \right]^{1/2}
\end{equation}
is finite, so we can use a Sobolev embedding to obtain
\begin{equation}
\max(||v_x||_{L^\infty(\Omega)},||v_t||_{L^\infty(\Omega)})< \infty.
\end{equation}
This inequality will prove itself very useful in order to estimate the functional
\begin{equation}
E_3(t)=\frac{1}{2}\int_\mathbb{R}(v_{ttt}^2+c^6 v_{xxx}^2)dx.
\end{equation}
Its first derivative reads
\begin{equation}
\label{e3p}
E_3'(t)=\int_\mathbb{R}(v_{ttt}v_{tttt}+c^6 v_{xxx}v_{xxxt})dx=\int_\mathbb{R}(v_{ttt}v_{tttt}-c^6 v_{xxxx}v_{xxt})dx.
\end{equation}
Let us start with the first integral in the right hand side of this equation
\begin{equation}
\int_\mathbb{R} v_{ttt}v_{tttt} dx= -\frac{1}{\mu}\int_\mathbb{R}v_{ttt}^2 dx
-\frac{1}{\mu}\int_\mathbb{R}v_{ttt}(vv_x)_{tt} dx +c^2\int_\mathbb{R}v_{ttt}v_{xxtt}=K_1+K_2+K_3.
\end{equation}
We see that $K_1 \le 0$,
\begin{eqnarray}
\nonumber
K_2=-\frac{1}{\mu}\int_{\mathbb{R}}[v_{ttt}(v_{tt}v_x+2v_tv_{xt}+vv_{xtt})] \le
\frac{||v_x||_{L^\infty(\Omega)}}{2\mu^2}\int_{\mathbb{R}}(\mu^2 v_{ttt}^2+v_{tt}^2)dx+ \\
\frac{||v_t||_{L^\infty(\Omega)}}{\mu^2 c}\int_{\mathbb{R}}(\mu^2 v_{ttt}^2+c^2 v_{xt}^2)dx+
\frac{||v||_{L^\infty(\Omega)}}{2 \mu c}\int_{\mathbb{R}}(v_{ttt}^2+c^2 v_{xtt}^2)dx,
\end{eqnarray}
and
\begin{equation}
K_3=-\int_\mathbb{R}v_{xttt}v_{xtt}dx=-\frac{1}{2}\frac{d}{dt}\int_\mathbb{R}v_{xtt}^2 dx.
\end{equation}
The second integral in the right hand side of Eq.(\ref{e3p}) may be evaluated as follows
\begin{eqnarray}
\nonumber
c^2\int_\mathbb{R}v_{xxx}v_{xxxt}dx=-c^2\int_\mathbb{R}v_{xxt}v_{xxxx}dx= \\
-\int_\mathbb{R}\{v_{xxt}[v_{xxtt}+\mu^{-1} v_{xxt}+\mu^{-1} (vv_x)_{xx}]\} dx=L_1+L_2+L_3.
\end{eqnarray}
We see that $L_2 \le 0$,
\begin{equation}
L_1=-\frac{1}{2}\frac{d}{dt}\int_\mathbb{R}v_{xxt}^2 dx,
\end{equation}
and
\begin{eqnarray}
\nonumber
L_3=-\frac{1}{\mu}\int_\mathbb{R}v_{xxt}(3v_xv_{xx}+vv_{xxx})dx \le
\frac{3||v_x||_{L^\infty(\Omega)}}{2 \mu^2}\int_\mathbb{R}(\mu^2 v_{xxt}^2+v_{xx}^2) dx \\
+ \frac{||v||_{L^\infty(\Omega)}}{2 \mu c}\int_\mathbb{R}(v_{xxt}^2+c^2 v_{xxx}^2)dx.
\end{eqnarray}
Collecting these results, we can transform Eq.(\ref{e3p}) into the inequality
\begin{eqnarray}
\nonumber
\frac{d}{dt}\int_\mathbb{R}(v_{ttt}^2+c^2 v_{xtt}^2+c^4 v_{xxt}^2+c^6 v_{xxx}^2)dx \le
D_1 \int_\mathbb{R}(v_{tt}^2+c^2 v_{xt}^2+c^4 v_{xx}^2)dx + \\
D_2 \int_\mathbb{R}(v_{ttt}^2+c^2 v_{xtt}^2+c^4 v_{xxt}^2+c^6 v_{xxx}^2)dx,
\end{eqnarray}
for suitable positive constants $D_1$, $D_2$. Now, integrating this equation with respect to time we obtain
\begin{eqnarray}
\nonumber
\left[ \int_\mathbb{R}(v_{ttt}^2+c^2 v_{xtt}^2+c^4 v_{xxt}^2+c^6 v_{xxx}^2)dx \right](T_2) \le \\
\tilde{D}_1 + \tilde{D}_2 \int_0^{T_2} \int_\mathbb{R}(v_{ttt}^2+c^2 v_{xtt}^2+c^4 v_{xxt}^2+c^6 v_{xxx}^2)dxdt,
\end{eqnarray}
for $0 \le T_2 \le T$, where we have used the boundedness of the Sobolev norm (\ref{sobnorm}). Let us recall once again Gr\"{o}nwall inequality to get
\begin{equation}
\left[ \int_\mathbb{R}(v_{ttt}^2+c^2 v_{xtt}^2+c^4 v_{xxt}^2+c^6 v_{xxx}^2)dx \right](T_2) \le \tilde{D}_1 \exp(\tilde{D}_2 T_2).
\end{equation}
Integrating this expression with respect to $T_2$ in the interval $[0,T]$ yields
\begin{equation}
\int_0^T \int_\mathbb{R}(v_{ttt}^2+c^2 v_{xtt}^2+c^4 v_{xxt}^2+c^6 v_{xxx}^2)dxdt < \infty,
\end{equation}
and now, if we put this result together with the boundedness of $||v||_{H^2(\Omega)}$ we obtain the boundedness of the norm $||v||_{H^3(\Omega)}$ which implies, after invoking the corresponding Sobolev embedding, the desired result
$v \in C^2(\Omega)$.
$\Box$

{\bf Corollary.} {\it Consider the Cauchy problem for Eq.(\ref{hburgers}) with initial conditions fulfilling the same properties as in Theorems 1 and 2. Then there is a finite time $t_m$ such that 
$\lim_{t \to t_m} ||v||_{L^\infty(\mathbb{R})}=\infty$.}

In summary, we have proven two results that were previously conjectured in the physics literature concerning the hyperbolic modification of the Burgers equation. The first one establishes the finite speed of propagation of disturbances in this type of flow, and provides us with an estimation of the velocity of sound within it, $c=\sqrt{\nu / \mu}$. The second one shows that the solution to the hyperbolic Burgers equation blows up in finite time, provided that the initial condition is large enough, a result that complements the partial regularity theorems proven by Rudjak and Smagulov~\cite{rudiak}. In contrast to what happens in the case of an inviscid compressible fluid~\cite{sideris}, the flow does not develop a shock-wave, but the velocity itself diverges in finite time. Physically, we have shown what effects have in the macroscopic description including memory in the corresponding microscopic model. Although the undesirable property of infinite speed of signal propagation is lost, now we find that the nonexistence of the solution complicates the use of hydrodynamic equations to describe the coarse-grained dynamics of the microscopic process. Furthermore, the fluid might self-accelarate till reaching an infinite velocity, a fact that is clearly nonphysical, and that rules out the possibility of using the hyperbolic Burgers equation for large initial conditions. We hope that the analysis of this model will facilitate the assessment of the more complex hydrodynamic equations that appear as asymptotic approximations of the Boltzmann equation when a regularized version of the Chapman-Enskog expansion is performed.

It is a pleasure to acknowledge Prof. Sir John Ball for encouraging discussions on this problem.
This work has been partially supported by the Ministerio de Educaci\'{o}n y Ciencia (Spain) through Projects No. EX2005-0976 and FIS2005-01729.

\end{document}